\documentclass{article}

\usepackage{svg}
\usepackage{textalpha}
\usepackage{csquotes}
\usepackage{url}

\usepackage[
backend=biber,
style=numeric,
sorting=none,
]{biblatex}

\usepackage[english]{babel}

\usepackage[letterpaper,top=2cm,bottom=2cm,left=3cm,right=3cm,marginparwidth=1.75cm]{geometry}

\usepackage{amsmath}
\usepackage{graphicx}
\usepackage[colorlinks=true, allcolors=blue]{hyperref}
\usepackage{authblk}
\usepackage{listings}

\title{State-Averaged Orbital-Optimized VQE: A quantum algorithm for the democratic description of ground and excited electronic states}

\author[1, 3]{Martin Beseda}
\author[1, 2, 3]{Silvie Illésová}
\author[4]{Saad Yalouz}
\author[1]{Bruno Senjean}

\affil[1]{ICGM, Université de Montpellier, CNRS, ENSCM, Montpellier, France}
\affil[2]{VSB - Technical University of Ostrava, 708 00 Ostrava, Czech Republic}
\affil[3]{IT4Innovations National Supercomputing Center, VSB - Technical University of Ostrava, 708 00 Ostrava, Czech Republic}
\affil[4]{Laboratoire de Chimie Quantique, Institut de Chimie, CNRS/Université de Strasbourg, 4 rue Blaise Pascal, 67000 Strasbourg, France}

\date{}

\begin{document}
\maketitle

\begin{abstract}
The electronic structure problem is one of the main problems in modern theoretical chemistry. While there are many
already-established methods both for the problem itself and its applications like semi-classical or quantum dynamics,
it remains a computationally demanding task, effectively limiting the size of solved problems. Fortunately, it seems,
that offloading some parts of the computation to Quantum Processing Units (QPUs) may offer significant speed-up,
often referred to as quantum supremacy or quantum advantage. Together with the potential advantage, this approach
simultaneously presents several problems, most notably naturally occurring quantum decoherence, hereafter denoted as
quantum noise and lack of large-scale quantum computers, making it necessary to focus on Noisy-Intermediate Scale
Quantum computers when developing algorithms aspiring to near-term applications. SA-OO-VQE package aims to answer both
these problems with its hybrid quantum-classical conception based on a typical Variational Quantum Eigensolver approach,
as only a part of the algorithm utilizes offload to QPUs and the rest is performed on a classical computer, thus
partially avoiding both quantum noise and the lack of quantum bits (qubits). The SA-OO-VQE has the ability to treat
degenerate (or quasi-degenerate) states on the same footing, thus avoiding known numerical optimization problems arising
in state-specific approaches around avoided crossings or conical intersections.
\end{abstract}

\section{Statement of Need}
\begin{figure}
    \centering
    \includegraphics[width=.35\textwidth]{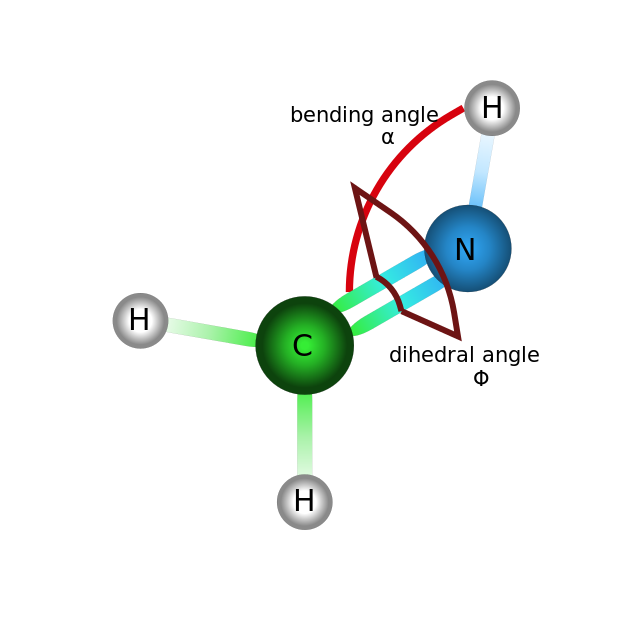}
    \caption{Molecule of formaldimine being described with bending and dihedral angles denoted $\alpha$ and $\phi$, respectively.}
    \label{fig:formaldimine}
\end{figure}
\begin{figure}
    \centering
    \includegraphics[width=.6\textwidth]{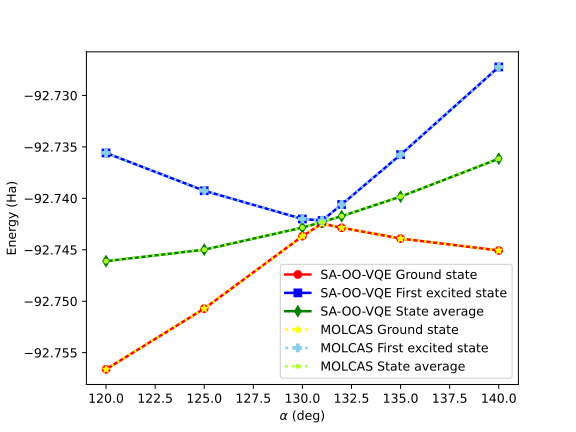}
    \caption{Comparison of potential energy depending on bending angle $\alpha$ in formaldimine molecule with dihedral angle 
$\phi = 90^\circ$.}
    \label{fig:energies}
\end{figure}
\begin{figure}
    \centering
    \includegraphics[width=.6\textwidth]{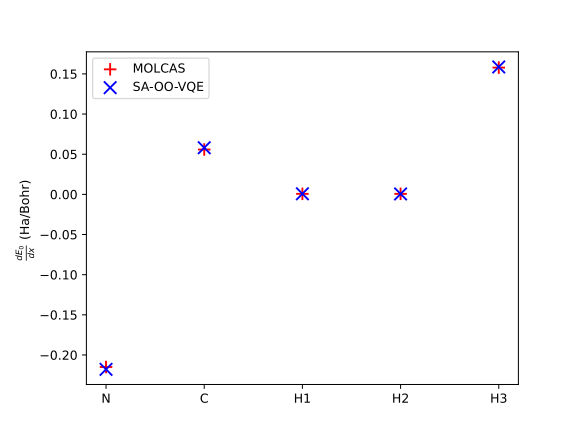}
    \caption{Comparison of ground-state gradients with bending angle $\alpha = 130^\circ$ and dihedral angle $\phi = 90^\circ$
in formaldimine molecule.}
    \label{fig:grad0}
\end{figure}
\begin{figure}[t]
    \centering
    \includegraphics[width=.6\textwidth]{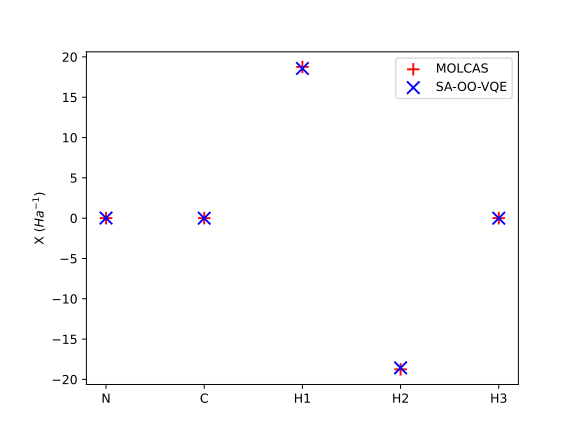}
    \caption{Comparison of total non-adiabatic couplings on bending angle $\alpha = 130^\circ$ and dihedral angle $\phi = 90^\circ$
in formaldimine molecule.}
    \label{fig:nac}
\end{figure}

Recently, quantum chemistry is one of the main areas-of-interest in Quantum Computing (QC)\cite{reiher2017elucidating,
mcardle2020quantum,bauer2020quantum}. That said, in many real-life applications, there is the necessity of treating
both the ground and excited states accurately and in an equal footing. The problem is magnified when the
Born-Oppenheimer approximation breaks down due to a strong coupling among degenerate or quasi-degenerate states,
most notably the ground and the first excited state, for which the accurate description requires (computationally demanding)
multi-configurational approaches. A good example of such case is a photoisomerization mechanism of the
rhodopsin chromophore, which progresses from the initial photoexcitation of the cis isomer over the relaxation in the first excited state
towards a conical intersection, where the population is transferred back to the ground state of the trans isomer. In order to describe such
a process thoroughly, one must compute not only relevant potential energy surfaces (PESs), but also their gradients
w.r.t. nuclear displacements, utilized further in molecular dynamics simulations. Finally, description of the conical
intersection can be done via obtaining non-adiabatic couplings (NACs).
Formally, the approaches describing PES topology, topography and the non-adiabatic couplings require Hamiltonian
diagonalization, which represents the most significant bottleneck. Considering classical methods like State-Averaged
Multi-Configurational Self-Consistent Field\cite{helgaker2013molecular}, only small complete active spaces have to be used
for the large computational overhead inherently present. However, such an approximation brings missing dynamical
correlation treatment, inducing the need to recover it ex-post, usually via some of the quasi-degenerate perturbation
techniques\cite{granovsky2011extended,park2019analytical}. On the other hand, QC brings the possibility of large complete
active spaces back, thus retaining a substantial part of dynamical correlation. Moreover, the dynamical correlation can
be also retrieved a posteriori utilizing QPUs only at the expense of more measurements, with no additional demands on
hardware infrastructure\cite{takeshita2020increasing}.
State-Averaged Orbital-Optimization Variational Quantum Eigensolver (SA-OO-VQE) method addresses the above-mentioned
problems and provides a way to compute both PES gradients and NACs analytically\cite{yalouz2021state,
yalouz2022analytical,omiya2022analytical}. Authored by Yalouz et al., there is an exemplary implementation\cite{saoovqe-old} focusing on
the pedagogical aspect and relying on matrix-vector multiplications rather than actual measurements, avoiding
the utilization of real QC infrastructure. Our implementation differs in a way that it aims to be a production-ready
solver utilizing both QCs and high-performance computing infrastructure in an efficient way, being able to run with
different backgrounds, utilizing Qiskit toolbox interface. The whole code is written in Python3, with YAML scripts
enabling its fast installation and usage.
The results are illustrated on molecule of formaldimine, which can be seen in \autoref{fig:formaldimine}.
Their comparison with the ones obtained via Molcas\cite{li2023openmolcas} implementation of
Complete Active-Space Self-Consistent Field\cite{malmqvist1989casscf} are shown in
\autoref{fig:energies}, \autoref{fig:grad0} and \autoref{fig:nac}. All the computations were computed with 3 active
orbitals containing 4 electrons and with STO-3G basis set.

\clearpage
\section{Features}
With SA-OO-VQE you can obtain following quantitites:

\begin{itemize}
    \item Potential energy surfaces
    \item Circuit (or Ansatz) gradients
    \item Orbital gradients
    \item Gradients of potential energy surfaces
    \item Non-adiabatic couplings
\end{itemize}

Also, for numerical optimization you can use any of the optimizers supported by Qiskit\footnote{\url{https://qiskit.org/documentation/stubs/qiskit.algorithms.optimizers.html}} and our own implementation of

\begin{itemize}
    \item Particle Swarm Optimization
\end{itemize}

\section{Getting Started}
The package is prepared with a priority of being very simple to use and the concise documentation can be found at \texttt{sa-oo-vqe-qiskit.rtfd.io}. To simplify the installation part, we recommend
utilizing Conda management system\footnote{\url{https://docs.conda.io/en/latest/}} together with prepared environment.yml file.
At first, users should clone the repository.

\begin{lstlisting}
git clone git@gitlab.com:MartinBeseda/sa-oo-vqe-qiskit.git
\end{lstlisting}

And install all the dependencies.

\begin{lstlisting}
$ conda env create -f environment.yml
$ conda init bash
$ source ~/.bashrc
$ conda activate saoovqe-env
$ pip install .
\end{lstlisting}

These commands run in a terminal will download and install all the necessary packages. The package availability can be
tested afterward simply by importing the package and looking at its version.

\begin{lstlisting}
$ python3
\end{lstlisting}

\begin{lstlisting}
>>> import saoovqe
>>> saoovqe.__version__
\end{lstlisting}
Finally, usage examples are located both in \texttt{examples} folder and in the documentation.

\section{Acknowledgements}
This work/project was publicly funded through ANR (the French National Research Agency) under
the "Investissements d’avenir" program with the reference ANR-16-IDEX-0006. This work was also
supported by the Ministry of Education, Youth and Sports of the Czech Republic through
the e-INFRA CZ (ID:90254).

\printbibliography

\end{document}